\documentclass[prl,twocolumn,aps]{revtex4-2}
\usepackage{amsmath}
\usepackage{amssymb}
\usepackage{amsthm}
\usepackage{graphicx}
\usepackage{bm}
\usepackage{physics}
\usepackage{mathtools}
\usepackage{xcolor}
\usepackage[caption=false]{subfig}

\usepackage{afterpage}

\usepackage{hyperref}
\definecolor{myblue}{RGB}{0,0,255}
\hypersetup{
colorlinks,
citecolor=myblue,
linkcolor=myblue,
urlcolor=myblue
}

\usepackage{subfig}
\usepackage{stmaryrd}
\usepackage{multirow}

\usepackage{booktabs}
\usepackage{float}

\begin{document}

\title{Thermodynamic Recycling of Algorithmic Failure Branches: Quantum-Computer Demonstration with Quantum Error Correction}
\author{Nobumasa Ishida}
\email{ishida@biom.t.u-tokyo.ac.jp}
\affiliation{Department of Information and Communication Engineering, Graduate School of Information Science and Technology, The University of Tokyo, Tokyo 113-8656, Japan}
\author{Yoshihiko Hasegawa}
\email{hasegawa@biom.t.u-tokyo.ac.jp}
\affiliation{Department of Information and Communication Engineering, Graduate School of Information Science and Technology, The University of Tokyo, Tokyo 113-8656, Japan}

\begin{abstract}
Thermodynamic trade-off relations dictate fundamental limits on the performance of thermodynamic tasks through costs such as heat dissipation. Here, we propose a framework called thermodynamic recycling to circumvent these limits in quantum processors by exploiting failure branches of quantum algorithms, which are usually discarded. The key component is an athermal bath naturally generated during the resetting of a failure branch. By coupling this bath to a target system prior to relaxation, thermodynamic tasks can be performed beyond conventional thermodynamic limits. We apply this framework to information erasure and derive the reduction in heat dissipation analytically. As a demonstration, we implement our framework on IBM's superconducting quantum processor by combining the Harrow--Hassidim--Lloyd algorithm with three-qubit quantum error correction, thereby reducing the heat dissipated in erasing syndrome information. Despite substantial noise and errors in current hardware, our method achieves erasure with heat dissipation below the Landauer limit. This work establishes an operational connection between quantum computing and quantum thermodynamics for resource-efficient quantum computation.
\end{abstract}

\maketitle
\textit{Introduction---}Quantum thermodynamics has revealed the necessity of precise resource management in cryogenic quantum systems. Thermodynamic trade-off relations show that the performance of thermodynamic processes is fundamentally bounded by available resources or, equivalently, thermodynamic costs such as heat dissipation \cite{Erker2017-hb,Campbell2017-wp,Brandner2018-cq,Chitambar2019-rx,Funo2019-cm,Carollo2019-lk,Hasegawa2020-rh,Van_Vu2022-hm,Hasegawa2023-de,Ishida2025-vt}. As quantum computers develop, such thermodynamic considerations are increasingly recognized within quantum computing \cite{Buffoni2022-jf,Bassman-Oftelie2024-nf,Blok2025-fh,Campbell2025-hk}: for example, in studies of heat engines \cite{Aamir2025-qq}, quantum batteries \cite{Anonymous2026-kd}, and quantum error correction (QEC) \cite{Simbierowicz2024-re,Krinner2019-ke,Bilokur2024-ck,Landi2020-cj,Danageozian2022-az}. In particular, QEC protocols require frequent erasure of syndrome information, each incurring heat dissipation dictated by Landauer's principle \cite{Landauer1961-gm}, which can become substantial when quantum computers scale to millions of qubits \cite{Bilokur2024-ck}.

Despite the scarcity of thermodynamic resources in cryogenic environments, there exists an overlooked yet prevalent resource: failure branches in quantum computing. Branch selection conditioned on measurements is ubiquitous in quantum algorithms. Examples range from the Harrow--Hassidim--Lloyd (HHL) algorithm for solving linear systems of equations~\cite{Harrow2009-ks} and quantum singular-value transformation~\cite{Gilyen2019-pa,Martyn2021-cw} to magic-state distillation~\cite{Bravyi2005-el,Bravyi2012-gi,Howard2017-ik}, a core subroutine of fault-tolerant quantum computing. In such protocols, only designated outcomes are retained, whereas the remaining outcomes correspond to failure branches. These branches are typically discarded and reset for subsequent trials. Despite efforts to reduce such discarding, including amplitude amplification and magic-state cultivation \cite{Grover1996-pr,Berry2014-dw,Gidney2024-nf}, failure branches themselves remain unavoidable and are considered wasteful byproducts.

In this Letter, we propose \textit{thermodynamic recycling}, a generic framework that converts failure branches into resources for thermodynamic tasks. Generally, when a failure branch is reset by being coupled to a cold finite bath, the bath is driven out of equilibrium. Our approach then couples this transient athermal bath to a target system before relaxation, enabling thermodynamic tasks to be performed beyond the conventional limits imposed by an equilibrium bath. While the framework applies to general branching algorithms and thermodynamic tasks, we analyze information erasure as a representative example. As a theoretical foundation, we derive the corresponding reduction in the minimum heat dissipation relative to equilibrium-bath bounds dictated by Landauer's principle \cite{Landauer1961-gm,Timpanaro2020-lr}. Then, we demonstrate the framework on IBM's superconducting quantum processor~\cite{IBMUnknown-du}, where we supply the failure branch of the HHL algorithm to a three-qubit QEC circuit via classical feedforward operation. As a result, we observe a reduction in heat dissipation in erasing syndrome information below Landauer's principle, even with the limitations of current hardware, such as short coherence time and significant noise. Our work suggests that thermodynamic recycling is a viable strategy for near-term quantum computers.

\begin{figure*}[ht]
\centering
\includegraphics[width=0.9\linewidth]{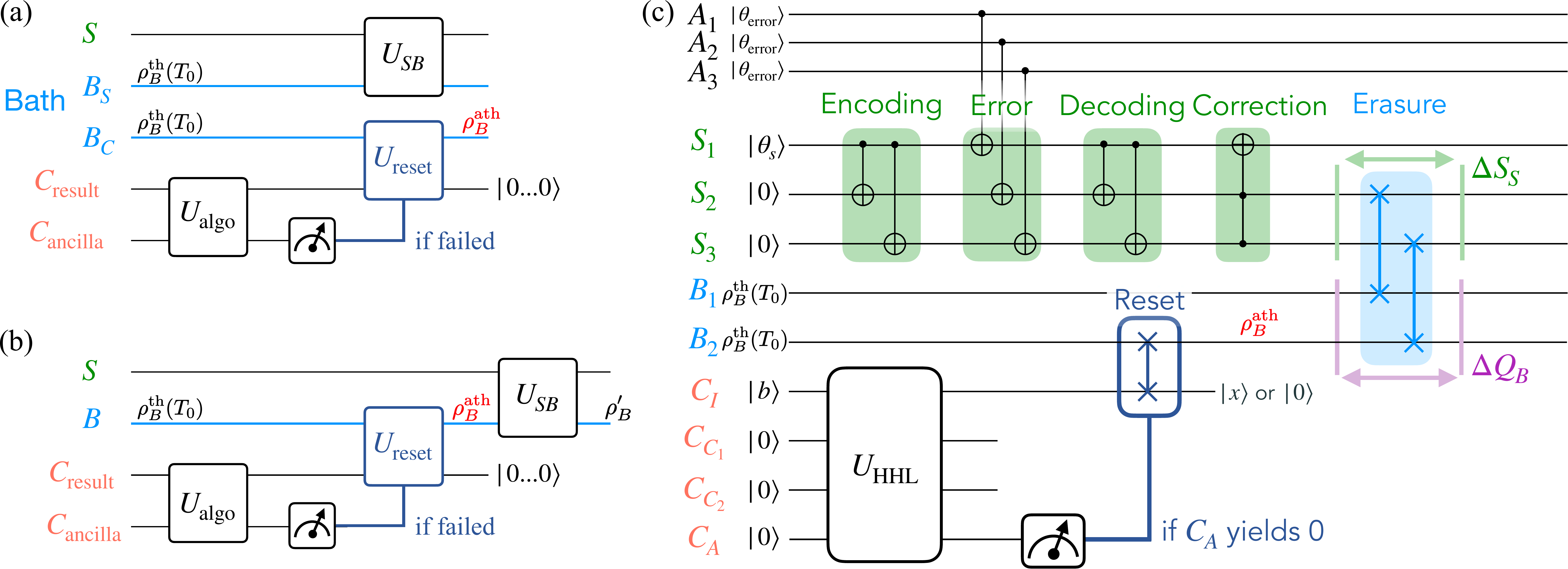}
\caption{Illustration of thermodynamic recycling. (a) A conventional setting where a quantum algorithm and a thermodynamic task are executed in parallel with separate baths. (b) Proposed setting where the two baths are unified, and the athermal bath from the failure branch is used for the thermodynamic task. (c) Circuit sketch of the demonstration. The failure branch of the HHL algorithm is reset by a SWAP operation with the bath $B$ via classical feedforward. Before $B$ relaxes, we implement the erasure of QEC ancillae by another SWAP between them and $B$: its performance is evaluated by the amount of entropy decrease $\Delta S_S$ and heat dissipation $\Delta Q_B$. The ancillary qubits $A_1A_2A_3$, initially in a superposition state $\ket{\theta_{\rm error}}$ depending on the error rate $p_x$, are used to apply an independent bit-flip channel to the encoded qubits $S_1S_2S_3$.}
\label{fig:scheme}
\end{figure*}

\textit{Setup---}We consider two processes executed on the same quantum-computing platform: a quantum algorithm on a register $C$ and a thermodynamic task on a target system $S$, as illustrated in Fig.~\ref{fig:scheme}(a). In the reference setting, the two processes use identical finite baths $B_{C}$ and $B_{S}$. For simplicity, we assume each bath is initially prepared in the same Gibbs state $\rho_B^{\mathrm{th}}(T_0)=e^{-\beta H_B}/Z$ at a low temperature $T_0$, where $H_B$ is the bath Hamiltonian, $\beta=(k_B T_0)^{-1}$, and $Z={\rm Tr}[e^{-\beta H_B}]$. A natural scenario for concurrent execution is quantum multi-programming~\cite{Das2019-gg,Dou2020-rg,Niu2023-ol,Niu2022-tt}, in which multiple programs run in parallel on a single processor to improve throughput on large-scale devices. In such an architecture, one program may include branch selection and therefore generate failure branches that must be reset, while another may require a thermodynamic task such as ancilla erasure in QEC.
 
The algorithm applies a unitary $U_{\rm algo}$ to $C$ and measures an ancillary subsystem $C_{\rm ancilla}$ to select the branch. The non-ancillary part is denoted by $C_{\rm result}$. In the success branch, $C_{\rm result}$ gives the desired output. In the failure branch, $C_{\rm result}$ must be reset for a retry. Standard reset procedures include measurement-based reset and bath-based reset \cite{Riste2012-bb,Krantz2019-dw,Zhou2021-ah,Bassman-Oftelie2024-nf,IBMUnknown-du}; here we focus on the latter for thermodynamic consistency. The coupling between $C_{\rm result}$ and the bath $B_C$ via a unitary $U_{\rm reset}$ yields
\begin{equation}
\rho_{\rm result}^{\rm fail}\otimes \rho_B^{\mathrm{th}}(T_0)
\xrightarrow{\ U_{\rm reset}\ }
\rho_{\rm result}^{\mathrm{init}}\otimes \rho_{B}^{\mathrm{ath}}.
\label{eq:reset}
\end{equation}
Here the failure branch returns to its initial state, while $B_C$ transitions to a nonequilibrium (athermal) state $\rho_{B}^{\mathrm{ath}}$. This athermality is pronounced when the bath is cold and has a few degrees of freedom, as is typical in quantum systems \cite{Esposito2010-eb,Reeb2014-jz,Uzdin2018-pr,Timpanaro2020-lr}. In the conventional setting, $B_C$ simply relaxes back to equilibrium and the athermality is wasted. Meanwhile, the thermodynamic task on $S$ uses the separate equilibrium bath $B_S$ via a unitary $U_{SB}$, and therefore its performance is subject to the standard thermodynamic bounds for an equilibrium bath. The induced map on $S$, i.e., a completely positive and trace-preserving map, can implement general thermodynamic tasks such as information erasure, heat-engine operation, or battery charging \cite{Vinjanampathy2016-mw,Reeb2014-jz,Goold2015-is,Rodrigues2019-ec,Landi2024-vn,Proesmans2017-hw,Hasegawa2025-ui}.

\textit{Thermodynamic recycling---}Our approach unifies the two baths, $B_S=B_C\equiv B$ as shown in Fig.~\ref{fig:scheme}(b). After the reset of the failure branch drives $B$ out of equilibrium, we immediately couple $B$ to the target system $S$ and perform $U_{SB}$ \textit{before} $B$ relaxes:
\begin{equation}
\rho_S\otimes \rho_B^{\mathrm{ath}}
\xrightarrow{\ U_{SB}\ } \rho'_{SB}, \quad
\rho'_S={\rm Tr}_B[\rho'_{SB}],\
\rho'_B={\rm Tr}_S[\rho'_{SB}].
\end{equation}
The athermality of the bath, including coherence and non-Gibbs energy distributions, can enhance the performance of thermodynamic tasks or, equivalently, reduce the thermodynamic cost beyond the conventional limits achievable with an equilibrium bath. Therefore, we refer to this process as thermodynamic recycling. This framework is generic: it does not depend on a specific algorithm or thermodynamic task.

Thermodynamic recycling is distinct from reservoir engineering \cite{Mari2012-ur,Rossnagel2014-ok,Konopik2020-up,Zhang2023-ke,Campbell2017-jk}, which also exploits nonequilibrium baths to enhance thermodynamic performance. In reservoir engineering, the athermal bath is actively prepared by an additional external driving stage on the bath, which incurs an extra cost. Considering the total workflow, including the bath preparation, the net performance cannot surpass the equilibrium-bath limit. In our approach, by contrast, the athermal bath is generated passively as a byproduct of the reset that is inevitable in quantum computation, and therefore incurs no additional cost.

Hereafter, we specialize in information erasure to clarify the effectiveness of thermodynamic recycling in reducing thermodynamic costs. Let $\Delta S_S=S(\rho_S)-S(\rho'_S)>0$ be the decrease of von Neumann entropy of the system $S$ under the process $U_{SB}$, and let $\Delta Q_B={\rm Tr}[H_B(\rho'_B-\rho_B)]$ be the heat transferred to the bath. If the bath is initially in equilibrium at temperature $T_0$, Landauer's principle gives the lower bound $\Delta Q_B \geq Q_{\rm Landauer}= k_B T_0 \Delta S_S$~\cite{Landauer1961-gm,Esposito2010-eb,Reeb2014-jz}. However, for the cold finite baths relevant here, the sharper equilibrium baseline is the finite-bath bound $\Delta Q_B \geq Q_{\rm tight}(\Delta S_S;T_0) \geq Q_{\rm Landauer}$, where $Q_{\rm tight}$ is characterized by the bath heat capacity (see End Matter for the explicit form) \cite{Timpanaro2020-lr}. In the following, we use $Q_{\rm tight}$, rather than $Q_{\rm Landauer}$, as the reference limit for an equilibrium bath; for sufficiently small and cold baths, $Q_{\rm Landauer}$ is generally not saturable and thus obscures the effect of thermodynamic recycling.

In thermodynamic recycling, the erasure step $U_{SB}$ acts on the athermal bath  $\rho_B=\rho_B^{\mathrm{ath}}$ produced by failure-branch reset. The corresponding lower bound on heat dissipation becomes the following:
\begin{equation}
Q_{\rm ath}(\Delta S_S;T_0,\rho_B^{\mathrm{ath}})= Q_{\rm tight}(\Delta S_S;T_0)-G(\Delta S_S;T_0,\rho_B^{\mathrm{ath}}),\label{eq:Q_ath}
\end{equation}
where the gain $G$ quantifies the reduction enabled by the athermality of the bath:
\begin{align}
&G(\Delta S_S;T_0,\rho_B^{\rm ath})\nonumber\\
&=\Delta E(\rho_B^{\rm ath}, \rho_B^{\rm th}(T_0))- k_B\int_{S(\rho_B^{\rm th}(T_0))}^{S(\rho_B^{\rm ath})}\mathcal{T}(s+\Delta S_S)ds.\label{eq:gain}
\end{align}
Here $\Delta E(\rho,\sigma)={\rm Tr}[H_B(\rho-\sigma)]$ is the energy difference between $\rho$ and $\sigma$, and $\mathcal{T}(s)  (>0)$ denotes the temperature of the Gibbs state of the bath with entropy $s$. The first term in $G$ is the energetic advantage carried by the athermal bath, whereas the second term is the penalty associated with the entropy accumulation in the bath during branch reset. When $G>0$, the erasure of entropy $\Delta S_S$ can be achieved with dissipation below $Q_{\rm tight}$; the actual performance $\Delta Q_B$ depends on the design of $U_{SB}$ and on empirical noise. Since $Q_{\rm tight}> Q_{\rm Landauer}$ in our regime, crossing $Q_{\rm tight}$ is the primary signature of improvement, while crossing $Q_{\rm Landauer}$ is a stronger but more challenging benchmark. The derivation of Eqs.~\eqref{eq:Q_ath} and \eqref{eq:gain} is given in End Matter, and the condition for positive gain is analyzed in detail there.

The gain is closely related to the specific structure of branches in an algorithm. In algorithms with many failure branches, such as magic-state distillation~\cite{Bravyi2005-el}, different failure outcomes generally generate different athermal bath states. We can show that retaining the branch information can enhance the gain. Consider distinguishing these branches with a classical index $m$ and retaining this branch information. The subsequent thermodynamic step can then be conditioned on $m$, allowing it to exploit the specific athermality of each branch and achieve an average gain $\bar{G}$. By contrast, if branch information is discarded, the bath is described by the mixed state on all branches, resulting in another gain $G_{\rm mix}$. We show in End Matter that $G_{\rm mix}\leq\bar{G}$: in particular, the difference $\bar{G}-G_{\rm mix}$ is determined by the Holevo quantity $\chi$ on the branch indices, which quantifies our knowledge on the branch \cite{Nielsen2010-tb}.

\textit{Demonstration---}We demonstrate thermodynamic recycling on IBM's quantum processor ibm\_kawasaki,  a 156-qubit superconducting ``Heron'' device (see Supplemental Material for specifications) \cite{IBMUnknown-du,IBMUnknown-iw,McKay2023-tj}. We combine the HHL algorithm, which supplies the failure branches, with a QEC protocol, whose ancilla erasure is the target thermodynamic task. Our aim is to reduce the heat dissipation during the erasure of QEC ancillae while successfully executing both programs on the same device.  The circuit is illustrated in Fig.~\ref{fig:scheme}(c). The total system consists of 12 qubits: four for the HHL register, six for the QEC register, and two for the bath. Although baths for qubit reset in current processors are external to the processor and not directly manipulable, we implement the bath on-chip as a minimal effective model shared by the two programs, which enables us to track the associated thermodynamic quantities \cite{Solfanelli2021-rc,Ishida2025-vt}. As with the other qubits, the bath qubits are initialized by natural relaxation.

The HHL algorithm solves a $2\times2$ linear system $Ax=b$ with $A=\bigl(\begin{smallmatrix} 2 & -1\\ -1 & 2\end{smallmatrix}\bigr)$ and $b=(\cos\theta_b\;\sin\theta_b)^T$, using a simplified circuit~\cite{Zheng2017-ob}. The HHL register consists of an input/output qubit $C_I$, two clock qubits $C_C$, and a rotation ancilla $C_A$. The failure-branch state depends on $\theta_b$, and hence so does the athermality generated in the bath. A classical feedforward operation~\cite{Baumer2024-rz,Baumer2024-hg} conditionally triggers a reset of $C_I$ by coupling it to the bath $B$ only when the algorithm fails, simultaneously generating the athermal bath $\rho_B^{\mathrm{ath}}$. In particular, the reset is implemented by a SWAP operation between $C_I$ and one of the bath qubits $B_1$ \cite{Bassman-Oftelie2024-nf}.

For the QEC part, we implement a three-qubit repetition code protecting a target qubit $S_1$. To test the correction capability, we apply an independent bit-flip channel $\mathcal{E}_{p_{\rm error}}(\rho)=(1-p_{\rm error})\rho+p_{\rm error}X\rho X$ to each encoded qubit, with $p_{\rm error}=0.15$, through interaction with the auxiliary qubits $A_1A_2A_3$. We adopt an operator error correction \cite{Kribs2005-ko,Kribs2006-ua,Clemens2006-pr,Kondo2013-bz,Heusen2024-bt,Crow2016-ng,Ercan2018-cq,Perlin2023-lm}, where decoding and correction are realized only with coherent operations and no syndrome measurement is performed. Let $\rho_{S_1}^{\rm init}$ and $\rho_{S_1}^{\rm corrected}$ denote the initial and corrected states of $S_1$, respectively, and define $F_{\rm corrected}=F(\rho_{S_1}^{\rm init},\rho_{S_1}^{\rm corrected})$ and $F_{\rm error}=F(\rho_{S_1}^{\rm init},\mathcal{E}_{p_{\rm error}}(\rho_{S_1}^{\rm init}))$, where $F(\rho,\sigma)=({\rm Tr}\sqrt{\sqrt{\rho}\sigma\sqrt{\rho}})^2$ is the fidelity. We consider QEC to be successful when $F_{\rm error}<F_{\rm corrected}$, which is verified empirically below.

After correction, the ancilla qubits $S_2$ and $S_3$ retain syndrome information and thus acquire increased entropy. For the next cycle of correction, this information must be erased by coupling them to the bath, which necessarily dissipates heat $\Delta Q_B$ into the bath. The erasure operation is implemented by SWAP operations between $S_2S_3$ and the bath qubits $B_1B_2$. For thermodynamic recycling, we perform this erasure immediately after the HHL failure-branch reset, which transforms the bath into the athermal state. The central question is whether this procedure can reduce the heat dissipation $\Delta Q_B$ below the equilibrium-bath baselines $Q_{\rm tight}$ and $Q_{\rm Landauer}$ for a given entropy decrease $\Delta S_S$ of $S_2S_3$.

This question is nontrivial on current hardware, which suffers from noise and short coherence times. In particular, classical feedforward is much slower than coherent gates. While one- and two-qubit gates are executed within tens of nanoseconds for each, the feedforward operation takes several microseconds \cite{IBMUnknown-du}. The resulting latency degrades the bath athermality before the erasure step due to natural relaxation and is therefore expected to reduce the gain. The demonstration directly tests whether thermodynamic recycling yields a significant advantage despite these device constraints.

To estimate the relevant quantities, we perform state tomography at several points in the protocol \cite{Smolin2012-iy}.  Using the reconstructed density matrices, we evaluate $\Delta S_S$ and $\Delta Q_B$ for the erasure step. In particular, we use the bath Hamiltonian $H_B$ adopting 5 GHz as the resonance frequency of the involved qubits \cite{Solfanelli2021-rc,Krantz2019-dw,Gao2021-jc,Ishida2025-vt}; the qualitative conclusions are unchanged within the range 3--7 GHz. Combined with vendor calibration data, this gives an effective initial bath temperature $T_0=31$ mK \cite{Solfanelli2021-rc,Buffoni2022-jf,Bassman-Oftelie2024-nf}. Each tomography basis uses 4,000 shots, and each configuration of $\theta_b$ is repeated ten times to evaluate fluctuation.

\begin{figure}[t]
\centering
\includegraphics[width=\linewidth]{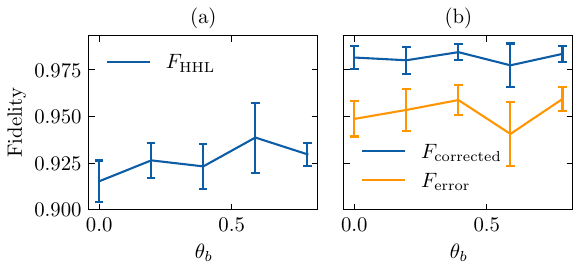}
\caption{Empirical fidelity of the HHL and QEC parts on ibm\_kawasaki. (a) HHL fidelity $F_{\rm HHL}(\theta_b)$ of the output state in the success branch. (b) QEC fidelities: the corrected state fidelity $F_{\rm corrected}$ (blue line) and the naive error fidelity $F_{\rm error}$ (orange line). Each line connects the mean values over ten trials, and the error bars represent the standard deviation over these trials.}
\label{fig:results_fidelity}
\end{figure}

We first empirically verify the functionality of the two constituent programs before discussing the erasure results. Figure~\ref{fig:results_fidelity}(a) shows the HHL fidelity $F_{\rm HHL}(\theta_b)=F(\ket{x(\theta_b)},\rho_{C_I}^{\rm Suc})$ where $\rho_{C_I}^{\rm Suc}$ is reconstructed by tomography. The fidelity remains around 0.92--0.93, indicating that the success branch yields the intended solution state \cite{Zheng2017-ob,Lee2019-dn}. The deviation from unity reflects substantial hardware noise of present-day devices. Furthermore, Fig~\ref{fig:results_fidelity}(b) shows $F_{\rm corrected}$ and $F_{\rm error}$. We find $F_{\rm error}<F_{\rm corrected}$ across the explored configurations, consistent with a successful correction. Theoretically, the QEC part is independent of $\theta_b$, and the data support this expectation. Together, these results confirm that the HHL and QEC programs operate properly in parallel.

\begin{figure*}[t]
\centering
\includegraphics[width=0.9\linewidth]{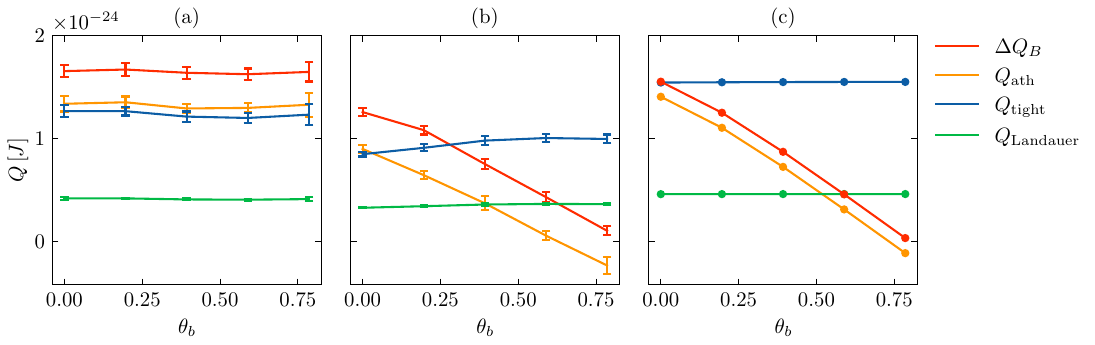}
\caption{Demonstration of thermodynamic recycling on ibm\_kawasaki. (a) Empirical heat dissipation $\Delta Q_B$ for the success branch, where the bath remains in equilibrium, is compared with the theoretical bounds $Q_{\rm ath}$, $Q_{\rm tight}$, and $Q_{\rm Landauer}$, for various $\theta_b$. (b) The same comparison is made for the failure branch, where thermodynamic recycling is implemented. (c) Theoretical prediction for thermodynamic recycling at $T_0=31$ mK.}
\label{fig:results_heat}
\end{figure*}

We next establish the equilibrium-bath reference by using the success branch, where the bath remains thermal at $T_0$ before erasure. Figure~\ref{fig:results_heat}(a) shows the measured $\Delta Q_B$ together with $Q_{\rm ath}$, $Q_{\rm tight}$, and $Q_{\rm Landauer}$, evaluated from the empirical $\Delta S_S$. The observed dissipation stays above all these bounds, consistent with the theory. The notable separation between $Q_{\rm tight}$ and $Q_{\rm Landauer}$ reflects the finite-size and low-temperature effect of the bath. In this reference setting, the theoretical expectation is $Q_{\rm ath}=Q_{\rm tight}$, and the data are consistent with this trend, except for a small discrepancy.

Our main result is shown in Fig.~\ref{fig:results_heat}(b), where the erasure of $S_2S_3$ is carried out using the athermal bath $\rho_B^{\mathrm{ath}}$ generated by HHL failure-branch reset. The data exhibit a clear regime in which the dissipated heat $\Delta Q_B$ lies below the conventional bound $Q_{\rm tight}$; for larger $\theta_b$, it also falls below the Landauer limit $Q_{\rm Landauer}$. Thus, for a given erased entropy $\Delta S_S$, thermodynamic recycling achieves a dissipation level that is inaccessible with an equilibrium bath. Equivalently, it reduces the net heat cost of erasure compared to performing HHL and erasure \textit{separately}. At the same time, all data remain above the nonequilibrium lower bound $Q_{\rm ath}$, in agreement with theory. We note that these theoretical bounds differ from those in Fig.~\ref{fig:results_heat}(a) because they depend on the empirical entropy decrease $\Delta S_S$, which differs substantially between the success-branch and failure-branch settings (see Fig.~\ref{fig:delta_S_s_success_failure} in End Matter).

Figure~\ref{fig:results_heat}(c) plots the theoretical prediction for thermodynamic recycling at $T_0=31$ mK. For sufficiently large $\theta_b$, the heat dissipation $\Delta Q_B$ is below both $Q_{\rm tight}$ and $Q_{\rm Landauer}$, consistent with the data. Even for smaller $\theta_b$, the theory predicts $\Delta Q_B<Q_{\rm tight}$, while the data show $\Delta Q_B>Q_{\rm tight}$. This discrepancy can be attributed to not only the empirical gap between $\Delta Q_B$ and the bound $Q_{\rm ath}$ but also the degradation of the gain $G$ from the theoretical prediction, as analyzed below.

\begin{figure}[t]
\centering
\includegraphics[width=0.9\linewidth]{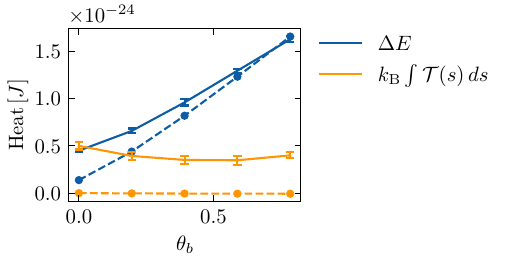}
\caption{Breakdown of the empirical gain $G$ into the energetic advantage $\Delta E(\rho_B^{\rm ath}, \rho_B^{\rm th}(T_0))$ and $k_B\int_{S(\rho_B^{\rm th}(T_0))}^{S(\rho_B^{\rm ath})}\mathcal{T}(s+\Delta S_S)ds$. The solid lines are empirical values, while the dashed lines are theoretical predictions. 
}
\label{fig:results_gain_terms}
\end{figure}
Figure~\ref{fig:results_gain_terms} shows the first and second terms in the gain $G$ in Eq.~\eqref{eq:gain} separately. The first term, representing the energetic advantage of the athermal bath, remains positive over the explored range of $\theta_b$ for both theory and experiment: the empirical values are even slightly larger than the theoretical ones. By contrast, the second term, representing the entropy-penalty contribution, is substantially larger in the demonstration than in theory. This indicates that the dominant degradation mechanism is not the loss of energy advantage but entropy increase resulting from decoherence during the HHL-reset-erasure sequence. In particular, not only the gate errors but also noise during the feedforward latency can contribute to this degradation. Despite these constraints, however, our demonstration shows that thermodynamic recycling yields a significant reduction of erasure cost on a current quantum processor.

\textit{Concluding remarks---}We proposed a framework that exploits failure branches, previously discarded in quantum computing, as a resource and demonstrated it on hardware. The significance of the demonstration is to simultaneously (i) achieve successful execution of both programs (HHL and QEC) and (ii) observe reduced heat dissipation below the equilibrium-bath bounds, regardless of hardware limitations. A promising direction for future work is to explore a broad class of computational subroutines and thermodynamic tasks to gain insights into the design of thermally efficient quantum computers.

\vspace{1em}
\begin{acknowledgments}
\noindent
\textit{Acknowledgments---}We acknowledge the use of IBM Quantum services for this work. The views expressed are those of the authors, and do not reflect the official policy or position of IBM or the IBM Quantum team.
\end{acknowledgments}

\newpage
\par
\appendix
\onecolumngrid
\begin{center}
\textbf{End Matter}
\end{center}
\twocolumngrid

\subsection{Derivation of the lower bound $Q_{\rm ath}$}
We derive the lower bound $Q_{\rm ath}$ in Eq.~\eqref{eq:Q_ath}. As a starting point, we clarify the tight bound $Q_{\rm tight}$ in the main text. Let the initial bath state be the thermal state at temperature $T_0$: $\rho_B=\rho_B^{\rm th}(T_0)$. The tight bound $Q_{\rm tight}(\Delta S_S;T_0)$ for erasing an entropy $\Delta S_S$ is given by \cite{Timpanaro2020-lr}:
\begin{align}
 Q_{\rm tight}(\Delta S_S;T_0)=\mathcal{Q}(\mathcal{S}^{-1}(\Delta S_S;T_0);T_0).\label{eq:Landauer-ZLP},
\end{align}
where the functions $\mathcal{S}$ and $\mathcal{Q}$ are defined as
\begin{align}
    \mathcal{S}(T;T_0)&=\int_{T_0}^{T}\frac{\mathcal{C}_B(\tau)}{k_B \tau}d\tau,\;\mathcal{Q}(T;T_0)=\int_{T_0}^{T}\mathcal{C}_B(\tau)d\tau.
\end{align}
Here, $\mathcal{C}_B(T)$ is the heat capacity of the bath at temperature $T$. The equality condition $\Delta Q_B=Q_{\rm tight}$ holds when the final bath state is a Gibbs state, and the final joint state of the system and bath is uncorrelated.

Next, we derive the bound $Q_{\rm ath}$ for an athermal bath $\rho_B^{\rm ath}$. Our derivation is based on the positivity of mutual information and the MaxEnt principle, based on Ref.~\cite{Timpanaro2020-lr}. The mutual information between $S$ and $B$ after the erasure process is $I_{SB} = S(\rho'_S) + S(\rho'_B) - S(\rho'_{SB}) \geq 0$. Using the fact that the initial entropy is given by $S(\rho_S) + S(\rho_B^{\rm ath})$ and that the von Neumann entropy is invariant under unitary evolution, we have $\Delta S_B \geq \Delta S_S$, where $\Delta S_B = S(\rho'_B) - S(\rho_B^{\rm ath})$ is the change in the bath entropy. Therefore, to erase an entropy $\Delta S_S$ from the system, the bath entropy must increase by at least $\Delta S_S$.

We consider the case where the athermal bath $\rho_B^{\rm ath}$ has the same entropy as the thermal state at temperature $T_0$: $S(\rho_B^{\rm ath}) = S(\rho_B^{\rm th}(T_0))$. Then, among the states with entropy at least $S(\rho_B^{\rm th}(T_0)) + \Delta S_S$, the Gibbs state at temperature $\mathcal{S}^{-1}(\Delta S_S; T_0)$ minimizes the energy due to the MaxEnt principle. In this case, we have
\begin{align}
    \Delta Q_B\geq Q_{\rm tight}(\Delta S_S;T_0)-\Delta E(\rho_B^{\rm ath},\rho_B^{\rm th}(T_0)),\label{eq:Landauer-athermal-equal-entropy}
\end{align}
where $\Delta E(\rho_B^{\rm ath},\rho_B^{\rm th}(T_0))$ is the energy difference between $\rho_B^{\rm ath}$ and $\rho_B^{\rm th}(T_0)$.

Now, we consider the general case where $S(\rho_B^{\rm ath})$ can differ from $S(\rho_B^{\rm th}(T_0))$. Let the difference be $r=S(\rho_B^{\rm ath}) - S(\rho_B^{\rm th}(T_0))$. When the bath entropy increases $\Delta S_S$, the final bath entropy becomes
\begin{equation}
S(\rho'_B) = S(\rho_B^{\rm ath}) + \Delta S_S = S(\rho_B^{\rm th}(T_0)) + \Delta S_S+ r.
\end{equation}
Therefore, the bath requires additional energy cost due to the extra entropy $r$ given by
\begin{align}
    \int_0^r{\mathcal{T}(s+\Delta S_S+S(\rho_B^{\rm th}(T_0)))ds=\int_{S(\rho_B^{\rm th}(T_0))}^{S(\rho_B^{\rm ath})}\mathcal{T}(s+\Delta S_S)ds,}
\end{align}
with $\mathcal{T}(s)=\pdv{E}{S}|_{S=s}$. Combining this with Eq.~\eqref{eq:Landauer-athermal-equal-entropy}, we obtain the lower bound for an athermal bath:
\begin{align}
    \Delta Q_B &\geq Q_{\rm tight}(\Delta S_S;T_0) - \Delta E(\rho_B^{\rm ath},\rho_B^{\rm th}(T_0)) \nonumber\\
    &\quad +k_B \int_{S(\rho_B^{\rm th}(T_0))}^{S(\rho_B^{\rm ath})}\mathcal{T}(s+\Delta S_S)ds.
\end{align}
This is the desired result in Eq.~\eqref{eq:Q_ath}.

\subsection{Analysis of the gain depending on the branch information}
The gain depends on the bath entropy: intuitively, the lower the bath entropy, the larger the gain. Here, we analyze how branch information affects the gain when multiple failure branches are present. Let each failure branch $m$ occur with probability $p_m$ and yield an athermal bath state $\rho_B^{\rm ath(m)}$ respectively. For simplicity, we assume that the bath entropies are equal for all $m$: $S(\rho_B^{\mathrm{ath}(m)})=S_{\rm failure}$. The gain for each branch is given by $G^{(m)}=G(\Delta S_S;T_0,\rho_B^{\rm ath(m)})$. The average gain reads 
\begin{align}
    &\bar{G}=\sum_{m=1}^M p_m G^{(m)}\nonumber\\
    &=\sum_{m=1}^M p_m \Delta E(\rho_B^{{\rm ath}(m)},\rho_B^{\rm th}(T_0))-k_B\int_{S(\rho_B^{\rm th}(T_0))}^{S_{\rm failure}}\mathcal{T}(s+\Delta S_S)ds.
\end{align}
As the energy difference is linear in the state, we have
\begin{align}
    \bar{G}=\Delta E(\bar{\rho}_B^{\rm ath},\rho_B^{\rm th}(T_0)) -k_B\int_{S(\rho_B^{\rm th}(T_0))}^{S_{\rm failure}}\mathcal{T}(s+\Delta S_S)ds,
\end{align}
where $\bar{\rho}_B^{\rm ath}=\sum_{m=1}^M p_m \rho_B^{\rm ath(m)}$ is the average bath state.

On the other hand, if we discard the branch information before erasure, the bath state becomes the mixture $\bar{\rho}_B^{\rm ath}$. Due to the concavity of the von Neumann entropy, the bath entropy increases as
\begin{align}
    S(\bar{\rho}_B^{\rm ath})=S(\sum_{m=1}^M p_m \rho_B^{{\rm ath}(m)})\geq \sum_{m=1}^M p_m S(\rho_B^{{\rm ath}(m)})=S_{\rm failure}.
\end{align}
Here, the increase is quantified by the Holevo quantity 
\begin{equation}
\chi=S(\bar{\rho}_B^{\rm ath})-S_{\rm failure}>0,
\end{equation}
which quantifies our knowledge of the branch index $m$ \cite{Nielsen2010-tb}. Therefore, the bath entropy is now $S(\bar{\rho}_B^{\rm ath})=S_{\rm failure}+\chi$. Using the quantity $\chi$, the gain for the mixed bath state is given by
\begin{align}
    G_{\rm mix}&= \Delta E(\bar{\rho}_B^{\rm ath},\rho_B^{\rm th}(T_0)) -k_B\int_{S(\rho_B^{\rm th}(T_0))}^{S(\bar{\rho}_B^{\rm ath})}\mathcal{T}(s+\Delta S_S)ds\nonumber\\
    &=\Delta E(\bar{\rho}_B^{\rm ath},\rho_B^{\rm th}(T_0)) -k_B\int_{S(\rho_B^{\rm th}(T_0))}^{S_{\rm failure}+\chi}\mathcal{T}(s+\Delta S_S)ds\nonumber\\
    &=\bar{G}-k_B\int_{S_{\rm failure}}^{S_{\rm failure}+\chi}\mathcal{T}(s+\Delta S_S)ds\nonumber\\
    &=\bar{G}-k_B\int_{0}^{\chi}\mathcal{T}(s+\Delta S_S+S_{\rm failure})ds.
\end{align}
As $\mathcal{T}>0$ and $\chi>0$, we have $G_{\rm mix}\leq \bar{G}$. Therefore, the more we know about the branch index $m$, the greater the gain we can obtain, and vice versa.

\subsection{Tradeoff between the erasure amount and the maximum tolerable bath entropy}

There exists a threshold $S_{\rm max}$ for the bath entropy such that a gain is obtained for $S(\rho_B^{\rm ath})<S_{\rm max}$: otherwise, no gain is obtained. We detail this theory here. Fix the bath energy difference $\Delta E>0$ and denote the athermal bath energy by $E^{\rm ath}={\rm Tr}[H_B \rho_B^{\rm ath}]$. We can show that there is a tradeoff between this threshold and the erasure amount $\Delta S_S$:
\begin{equation}
\pdv{S_{\rm max}}{\Delta S_S}<0.\label{eq:tradeoff_derivative}
\end{equation}
Therefore, the smaller the erasure amount $\Delta S_S$, the larger the bath entropy that can be tolerated while still yielding a gain, and vice versa. Moreover, consider the limit of infinitesimal erasure amount $\Delta S_S\to +0$. In this limit, we can show the following tradeoff relation:
\begin{equation}
\frac{S_{E^{\rm ath}}-S_{\rm max}}{\Delta S_S}\geq 1-\frac{T_0}{T_{E^{\rm ath}}},\label{eq:tradeoff}
\end{equation}
where $S_{E^{\rm ath}}$ is the entropy of the Gibbs state with energy $E^{\rm ath}$ and $T_{E^{\rm ath}}$ is its temperature. By the MaxEnt principle, $S_{E^{\rm ath}}\geq S_{\rm max}$. For $\Delta E>0$, $T_{E^{\rm ath}}>T_0$ holds in typical systems, so the right-hand side is positive. Therefore, Eq.~\eqref{eq:tradeoff} captures the universality of the gain. That is, for any $\rho_B^{\rm ath}$, unless it is a perfect equilibrium state, $S_{E^{\rm ath}}>S_{\rm max}$ holds, and for sufficiently small $\Delta S_S$ a gain is obtained.

We now derive Eqs.~\eqref{eq:tradeoff_derivative} and \eqref{eq:tradeoff}. By solving the equation $G(\Delta S_S;T_0,\rho_B^{\rm ath})=0$ for $S(\rho_B^{\rm ath})$, we obtain the threshold $S_{\rm max}$ as:
\begin{align}
    &S_{\rm max}(E_h)\nonumber\\&=\mathcal{S}(\mathcal{Q}^{-1}(E^{\rm ath}-E_B[\rho_B^{\rm th}(T_0)];T');T')+S(\rho_B^{\rm th}(T_0)),\label{eq:Smax-definition}
\end{align}
where
\begin{align}
    T'=\mathcal{S}^{-1}(-\Delta S_S;T_0).
\end{align}

To calculate the derivative of $S_{\rm max}$ with respect to $\Delta S_S$, we use the implicit function theorem. We have
\begin{align}
    \pdv{S_{\rm max}}{\Delta S_S}=\frac{\mathcal{T}(S(\rho_B^{\rm th}(T_0)-\Delta S_S)}{\mathcal{T}(S_{\rm max}-\Delta S_S)}-1.\label{eq:derivative-Smax-and-DeltaSs-tradeoff}
\end{align}
Under the weak assumption of $\mathcal{T}(a)>\mathcal{T}(b)$ for $a>b$, we have $\pdv{S_{\rm max}}{\Delta S_S}\leq 0$, which is Eq.~\eqref{eq:tradeoff_derivative}. Then, we obtain Eq.~\eqref{eq:tradeoff} by taking the limit $\Delta S_S\to +0$ in Eq.~\eqref{eq:derivative-Smax-and-DeltaSs-tradeoff} and using the Taylor expansion.

\subsection{Empirical entropy decrease $\Delta S_S$}
\begin{figure}[h]
\centering
\includegraphics[width=0.6\linewidth]{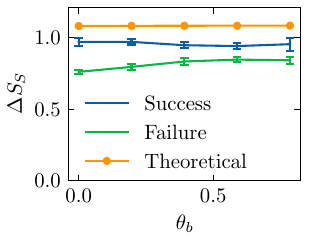}
\caption{Empirical entropy decrease $\Delta S_S$ of the system $S_2S_3$  for the success and failure branches, compared with the theoretical prediction.
}
\label{fig:delta_S_s_success_failure}
\end{figure}
Figure~\ref{fig:delta_S_s_success_failure} shows the empirical entropy decrease $\Delta S_S$ of the system $S_2S_3$ for the success and failure branches, compared with the theoretical prediction. The data show that $\Delta S_S$ is substantially smaller for the failure branch than for the success branch and theoretical prediction, which is the main reason for the discrepancy between the Y-axis values in Figs.~\ref{fig:results_heat}(a)--(c).

\end{document}